\documentclass[aps,preprint,groupedaddress,showpacs,nofootinbib]{revtex4}
\usepackage{graphicx}
\usepackage{ref}
\begin{document}
\title{A quark model analysis of the charge symmetry breaking in nuclear force}
\author{Takashi Nasu}
\author{Makoto Oka}
\affiliation{Department of Physics, Tokyo Institute of Technology, 
Meguro, Tokyo 152-8551,Japan}

\author{Sachiko Takeuchi}
\affiliation{Japan College of Social Work, Kiyose 204-8555, Japan}

\date{\today}

\begin{abstract}
In order to investigate the charge symmetry breaking (CSB) in the short range 
part of the nuclear force, we calculate the difference of the masses of the neutron and the proton, $\Delta {\rm M}$, 
the difference of the scattering lengths of the p-p and n-n scatterings, 
$\Delta a$, and the difference of the analyzing power of the proton 
and the neutron in the n-p scattering, $\Delta A(\theta)$, by a quark model. 
In the present model the sources of CSB are the mass difference 
of the up and down quarks and the electromagnetic interaction. 
We investigate how much each of them contributes to $\Delta {\rm M}$, 
$\Delta a$ and $\Delta A(\theta)$. It is found that the contribution of CSB 
of the short range part in the nuclear force 
is large enough to explain the observed $\Delta A(\theta)$, 
while $\Delta a$ is rather underestimated. 
\end{abstract}


\maketitle

\section{\label{sec1}Introduction}
The charge symmetry is the invariance under the charge-reflection,
{\it i.e.}, the reflection about the 1-2 plane in the isospin space. 
If this were an exact symmetry, 
the masses of the proton and the neutron would be the same, as well as 
the binding energies of the mirror nuclei or the scattering lengths of 
the p-p and n-n scatterings. 
The charge symmetry holds only approximately in the real world.
There are small but 
non-zero differences such as 
\begin{eqnarray}
\Delta {\rm M} = {\rm M}_n - {\rm M}_p =1.29~[{\rm MeV}]~~~~{\rm and}~~~~~~
\Delta a = a_{pp} -a_{nn} = 1.5 ~[{\rm fm}]~.
\end{eqnarray}
These differences are manifestation of the charge symmetry breaking 
(CSB). \par
CSB appears also in spin-dependent observables. 
For example, the ${\rm \vec{p}}$-n system is the mirror of 
${\rm \vec{n}}$-p, where ${\rm \vec{p}(\vec{n})}$ is a 
polarized nucleon. 
There was found small difference in the analyzing powers of 
${\rm \vec{p}}$ and ${\rm \vec{n}}$ in the medium energy scattering 
\cite{ALSex1,ALSex2}, 
\begin{eqnarray}
\Delta A(\theta)=A_n (\theta)-A_p (\theta)~. 
\end{eqnarray}
The study of $\Delta A(\theta)$ is 
important because there is no Coulomb interaction between 
n and p. 
\par
It is important to understand CSB from the quantum chromodynamics 
(QCD) viewpoint \cite{review}.
%
%
%
From QCD we find that
CSB has two origins: (i) the difference of the masses of the up 
and down quarks and (ii) the electromagnetic interaction. 
Thus the study of CSB phenomena can be a good probe to examine  
the behavior of the quarks and gluons in the low-energy region. 
The ultimate goal of the CSB study may be understanding their effects 
on hadron spectra and hadronic interactions directly from QCD, 
by, {\it e.g.}, lattice QCD simulation. 
As the direct approach is not available up to now, however,
indirect approaches have been taken for the CSB study. \par
An often used approach to CSB is based on the meson exchange picture 
of the nuclear force. 
It was suggested that CSB of the nuclear force is generated by 
mixings of $I=0$ and $I=1$ mesons such as $\rho$-$\omega$ 
mixing \cite{rhoomega1}. 
A model based on such a picture was reported to 
explain $\Delta a$ well.\cite{rhoomega2}
%
But it was also pointed out that the effect of the 
$\rho$-$\omega$ mixing to 
CSB may be suppressed by 
the off-shell effect of the $\rho$-$\omega$ mixing. \cite{offshell}
Thus, this problem is still open \cite{offshell2}. 
%
%
A class IV interaction \cite{text} is also generated by the neutron-proton 
mass difference in the one-pion-exchange interaction.\cite{ALSmeson} 
It was pointed out that the effects of OPE and $\rho$-$\omega$ 
mixing explain $\Delta A(\theta)$ fairly well. 
\par
On the other hand, 
CSB appearing in the short-range part should be investigated
by introducing subnucleonic degrees of freedom.
One of the pioneering works to apply a quark model
to CSB is found in Ref.~\onlinecite{Nakamura},
where the isovector mass shifts of isospin multiplets 
and the isospin-mixing matrix elements in 1s0d-shell nuclei are
investigated by using the quark cluster model (QCM) \cite{oka,fae,oy,shi,osysup}.
It was concluded that the u-d quark constituent mass difference 
produces significant effects, 
which may explain the observed Okamoto-Nolen-Schiffer anomaly \cite{ONS} 
well.

In the present work, we investigate
CSB in $\Delta {\rm M}$, $\Delta a$ and $\Delta A(\theta)$
by employing essentially the same model for all these three observables:
a quark potential model for $\Delta {\rm M}$
and QCM for $\Delta a$ and $\Delta A(\theta)$.
The CSB sources are taken to be 
(a) the difference of the masses of the up and down constituent quarks 
and (b) the electromagnetic interaction between the constituent quarks.
Our aim is to estimate the effect of CSB sources (a) and (b) 
on nuclear force by investigating the above three observables
simultaneously. 

Chemtob and Yang \cite{Yang} (CY) calculated $\Delta a$ using QCM, 
suggesting that the quark mass difference contributes to 
$\Delta a$ significantly. 
Later, Br\"{a}uer et al.\cite{brauer1,brauer2} studied $\Delta a$ and 
$\Delta A(\theta)$ using QCM and concluded 
that the effects of CSB sources (a) and (b) are too small 
to explain the observed value. 
However, their calculation of $\Delta A(\theta)$ 
suffers from a wrongly chosen factor, 
from omitting the symmetric spin-orbit term and from 
inconsistent use of the operators and wave functions (See sec~\ref{sec4}). 
\par 
%
%
%
In the present paper, we extend CY's and Br\"{a}uer's works in order to 
obtain more integrated knowledge on CSB. 
%
%
We investigate CSB in $\Delta {\rm M}$, $\Delta a$ and 
$\Delta A(\theta)$ simultaneously.
Also, we introduce the Instanton Induced Interaction (III) 
\cite{OT91,TO91,Ta94,Ta96,tH76,Sh84,Ko85}, 
which comes from the nonperturbative effects of QCD and explains the 
$\eta - \eta'$ mass splitting . 
Since III does not break 
the charge symmetry, its role in this study is mainly
to make the effective strength of the one-gluon exchange 
interaction smaller. The strength becomes reasonably small, 
which fits to the picture that this term represents the 
perturbative effect of the gluons (See sec \ref{sec4}). 
Moreover, we include the symmetric spin-orbit term in the 
analysis of $\Delta A(\theta)$, whose effect 
is as large as the antisymmetric
one. Furthermore,
we solve QCM to obtain the relative wave function
and use it to evaluate the matrix elements 
of $\Delta a$ and $\Delta A(\theta)$.

\par
In section \ref{sec2}, we show the Hamiltonian for quarks 
and the CSB sources. 
In section \ref{sec3}, we explain the detail of the calculations 
of $\Delta {\rm M}$, $\Delta a$ and $\Delta A(\theta)$. 
Results are discussed in section \ref{sec4}. 
Summary is given in section \ref{sec5}.
\section{\label{sec2}Hamiltonian} 
We employ the constituent quark model with quark masses of order 
$m \simeq 300$[MeV] in this study. 
The Hamiltonian is given by 
\begin{eqnarray}
{\rm H}&=& {\rm K}+{\rm V} 
\end{eqnarray}
K is the quark kinetic energy and considered as semirelativistic in 
calculation of $\Delta {\rm M}$ (See sec \ref{sec3.1}) 
and as non-relativistic in calculations of $\Delta a$ and $\Delta A(\theta)$ 
(See sec \ref{sec3.2}) in this study.
The quark-quark interactions are represented by a static potential, 
which consists of the confinement (CF), the one-gluon-exchange (OGE) 
\cite{CDGWZ78}, 
the electromagnetic (EM) and the instanton induced (III) interactions. 
\begin{eqnarray}
{\rm V}&=&{\rm V_{conf}}+{\rm V_{OGE}}+{\rm V_{EM}}
+{\rm V_{III}} \label{eq:V} \\
{\rm V_{CF}} &=& \sum_{i<j} -a(\vec{\lambda}_i \cdot \vec{\lambda}_j) 
r_{ij} \\
{\rm V_{OGE}} &=& \sum_{i<j} (\vec{\lambda}_i \cdot \vec{\lambda}_j )
\frac{\alpha_s}{4} \{~ \frac{1}{r_{ij}} - 
( \frac{\pi}{2m_i ^{2}}+\frac{\pi}{2m_j ^{2}}
+\frac{2\pi}{3m_i m_j}\vec{\sigma}_i \cdot \vec{\sigma}_j )
\delta (\vec{r}_{ij})  \nonumber \\
&-&[\frac{1}{2r^3_{ij}}(\frac{1}{m_i^2}+\frac{1}{m_j^2}+\frac{4}{m_i m_j})
]\vec{{\rm L}}_{ij} \cdot \frac{\vec{\sigma}_i + \vec{\sigma}_j}{2} \nonumber \\&-& [\frac{1}{4r^3_{ij}}(\frac{1}{m_i^2}-\frac{1}{m_j^2})
]\vec{{\rm L}}_{ij} \cdot \frac{\vec{\sigma}_i - \vec{\sigma}_j}{2}
\} \label{eq:LSOGE} \\
{\rm V_{EM}} &=& \sum_{i<j} e_i e_j
\alpha_{em} \{~ \frac{1}{r_{ij}} - 
(\frac{\pi}{2m_i ^{2}}+\frac{\pi}{2m_j ^{2}}
+\frac{2\pi}{3m_i m_j}\vec{\sigma}_i \cdot \vec{\sigma}_j)
\delta (\vec{r}_{ij}) \nonumber \\
&-&[\frac{1}{2r^3_{ij}}(\frac{1}{m_i^2}+\frac{1}{m_j^2}+\frac{4}{m_i m_j})]
\vec{{\rm L}}_{ij} \cdot \frac{\vec{\sigma}_i + \vec{\sigma}_j}{2} \nonumber \\
&-&[\frac{1}{4r^3_{ij}}(\frac{1}{m_i^2}-\frac{1}{m_j^2})
]\vec{{\rm L}}_{ij} \cdot \frac{\vec{\sigma}_i - \vec{\sigma}_j}{2}  
\} \label{eq:LSEM}
\\
%
%
%
{\rm V_{III}} &=& V_0^{(2)} \sum_{i<j} 
\Bigl(1+{3\over 32}\vec{\lambda}_i \cdot \vec{\lambda}_j+
{9\over 32}\vec{\lambda}_i \cdot \vec{\lambda}_j \vec{\sigma}_i 
\cdot \vec{\sigma}_j \Bigr)
\delta(\vec{r}_{ij})
 \nonumber \\
&-&
{1\over 8} \Bigl\{\Bigl(-1+{3\over 16}\lambda_i\cdot\lambda_j\Bigr)
\frac{2}{\bar{m}^2}
+{9\over 8\bar{m}^2}\lambda_i\cdot\lambda_j
\Bigr\}
{\delta(\vec{r}_{ij}) \over r^2}
\vec{{\rm L}}_{ij} \cdot \frac{\vec{\sigma}_i + \vec{\sigma}_j}{2}
%
%
%
%
%
\end{eqnarray}
$\vec{\lambda_i}$ is the color SU(3) Gell-Mann matrix and $e_i$ is the 
quark electric charge in units of the proton charge $e$. 
In this study it is assumed that the confinement potential does not break 
the charge symmetry. 
This is a natural assumption based on the confining potential obtained, 
for instance, from lattice QCD calculation. 
Yet there may exist velocity 
dependent terms associated with confinement which break the charge symmetry. 
We do not consider such terms in this study. 
Taking the Breit-Fermi interaction naively, non-Galilei invariant terms 
appear in the LS terms. 
But we consider only the Galilei invariant terms such as 
LS term in Eqs.~(\ref{eq:LSOGE}-\ref{eq:LSEM}). 
It should be noted that the Instanton Induced Interaction (III) 
is effective only 
on the flavor singlet (iso-singlet) quark-quark state. 
In other words, it works only 
on a pair of up and down quarks. 
Thus III does not break the charge symmetry. \par
In this Hamiltonian the terms including the quark mass and the electric charge 
may break the charge symmetry. In order to show the CSB terms explicitly
we rewrite the quark mass and the electric charge in terms of the 
isospin operator. 
\begin{eqnarray}
m_i &=& \frac{m_d+m_u}{2} - \frac{m_d-m_u}{2}\tau_3^{(i)} \nonumber \\
&=& {\bar m}(1-\frac{\Delta m}{2{\bar m}}\tau_3^{(i)}) \nonumber \\
&=& {\bar m}(1-\epsilon \tau_3^{(i)}) \\
e_i&=&\frac{\tau_{3}^{(i)}}{2}+\frac{1}{6} 
%
%
%
%
\end{eqnarray}
where 
\begin{eqnarray}
\bar{m}&=&\frac{m_d+m_u}{2} ~~~~~~~~~~ \Delta m = m_d - m_u \nonumber \\
\epsilon &=& \frac{\Delta m}{2\bar{m}}
\end{eqnarray}
Using the typical constituent quark mass 
${\bar m} \simeq 300$ MeV and the up and down quark mass difference 
$\Delta m \simeq 6$ MeV, 
$\epsilon \simeq \frac{6}{2\times 300}=\frac{1}{100}$ 
is as small as the electromagnetic coupling constant, 
$\alpha_{e.m.} \simeq 1/137$. 
So we divide the Hamiltonian into 
the charge symmetric part $\bar{{\rm H}}$ and the charge symmetry 
breaking part $\Delta {\rm H}_{{\rm CSB}}$, 
and treat ${\rm \Delta H_{CSB}}$ perturbatively. \par
The CSB part of the Hamiltonian is given to the leading order in 
$\epsilon$ and $\alpha_{e.m.}$ by 
\begin{eqnarray}
\Delta {\rm V_{CSB}} &=& \Delta {\rm V_{CSB}^{OGE}} 
+ \Delta {\rm V_{CSB}^{EM}} \label{eq:Delta V} \\
\Delta {\rm V_{CSB}^{OGE}} &=& \sum_{i<j} (\vec{\lambda}_i \cdot \vec{\lambda}_j )
\frac{\alpha_s}{4} \epsilon \{~ - 
\frac{\pi}{\bar{m}^{2}}(\tau_3^{(i)}+\tau_3^{(j)})
(1+\frac{2}{3}\vec{\sigma}_i \cdot \vec{\sigma}_j )
\delta (\vec{r}_{ij})  \nonumber \\
&-& \frac{3\alpha_s}{4\bar{m}^2 r_{ij}^3}{\rm \vec{L}_{ij}} 
\cdot (\vec{\sigma}_i+\vec{\sigma}_j)(\tau_3^{(i)}+\tau_3^{(j)}) \nonumber \\
&-& \frac{\alpha_s}{4\bar{m}^2 r_{ij}^3}{\rm \vec{L}_{ij}} 
\cdot (\vec{\sigma}_i-\vec{\sigma}_j)(\tau_3^{(i)}-\tau_3^{(j)})
\} \\
\Delta {\rm V_{CSB}^{EM}} &=& \sum_{i<j} \frac{\tau_3^{(i)}+\tau_3^{(j)}}{12}
\alpha_{e.m.} \{~ \frac{1}{r_{ij}} - \frac{\pi}{\bar{m}^2}
(1+\frac{2}{3}\vec{\sigma}_i \cdot \vec{\sigma}_j)
\delta (\vec{r}_{ij}) \nonumber \\
&-& \frac{3}{4\bar{m}^2 r_{ij}^3}{\rm \vec{L}_{ij}} 
\cdot (\vec{\sigma}_i+\vec{\sigma}_j) \} 
\end{eqnarray}
We ignore the second order terms ${\mathcal O}(\epsilon^2,\alpha_{e.m.}^2,\epsilon \alpha_{e.m.})$. 
The CSB terms from the tensor interaction are excluded because the tensor interactions between quarks are small. 
But we consider them when solving the charge symmetric equation 
for the unperturbated wave function. \par
%
%
The Hamiltonian has 5 parameters, $\alpha_s$,$\bar{m}$,$a$,$V_0^{(2)}$ and 
$\Delta m$. The parameters are determined so as to reproduce the 
single baryon properties and the results are shown in section \ref{sec4}. \\
\section{\label{sec3}Calculations}
In this section we present the formulas of the neutron-proton 
mass difference, $\Delta {\rm M}$, the difference of the scattering lengths 
of the p-p and n-n scattering, $\Delta a$, and the difference of 
the analyzing power of the neutron and the proton of the n-p scattering, 
$\Delta A(\theta)$. 
\subsection{\label{sec3.1}The proton-neutron mass difference $\Delta {\rm M}$}
The differences of the mass of the isodoublet hadrons were 
evaluated in the constituent quark model by Isgur \cite{Isgur}. 
We also evaluate the neutron-proton mass difference in order to 
determine the mass difference of the up and down constituent quarks. 
Our approach is different in the following two points. 
First, we consider the semi-relativistic kinetic energy term, 
%
\begin{eqnarray}
{\rm K} &=& \sum_i^3 \sqrt{m_i^2+p_i^2} \label{eq:semirel}
\end{eqnarray}
Eq.~(\ref{eq:semirel}) can be divided into the charge symmetric part 
and the charge symmetry breaking part, 
\begin{eqnarray}
{\rm K} &=& \bar{{\rm K}}+\Delta {\rm K_{CSB}} \\
\bar{{\rm K}} &=& \sum_i^3 \sqrt{\bar{m}^2+p_i^2} \\
\Delta {\rm K_{CSB}} &=& -\sum_i^3 \frac{\bar{m}^2}{\sqrt{\bar{m}^2+p_i^2}}
\epsilon \tau_3^{(i)} \label{eq:Delta K}
\end{eqnarray}
Eq.~(\ref{eq:semirel}) contains the kinetic energy of the center of mass coordinate, 
which must be subtracted in order to calculate the baryon mass. 
For the semirelativistic kinematics, the center-of-mass energy can not
be treated exactly.
Therefore we use the following approximation, 
\begin{eqnarray}
{\rm M_N} &=& 
\langle \sqrt{ {\rm H}^2 - {\rm P_G^2} } 
\rangle \nonumber \\
&\simeq& \langle {\rm H} \rangle 
- \frac{ \langle {\rm P_G^2} \rangle }{2\langle  {\rm H} \rangle } 
\end{eqnarray}
The relativistic effect is partially included as 
the convergence of the expansion in 
${\rm \frac{\langle P_G \rangle}{\langle H \rangle}}$ is 
better than that in $\langle \frac{p_i}{m_i} \rangle $. 
Then the nucleon mass can be written in terms of $\bar{\rm{H}}$ and 
$\Delta {\rm H}_{CSB}$ as 
\begin{eqnarray}
{\rm M_N} &=& \langle {\rm \bar{H}} \rangle - 
{\rm \frac{\langle P_G^2 \rangle}{2\langle \bar{H} \rangle} 
+\langle \Delta {\rm H_{CSB}} \rangle(1+\frac{\langle P_G^2 \rangle}
{2\langle \bar{H} \rangle ^2})} \label{eq:massformula}
\end{eqnarray}
where
\begin{eqnarray}
{\rm H} &=& \bar{{\rm H}} + \Delta {\rm H_{CSB}} \\
\bar{{\rm H}} &=& \bar{{\rm K}} + \bar{{\rm V}} \\
\Delta {\rm H_{CSB}} &=& \Delta {\rm K_{CSB}} + \Delta {\rm V_{CSB}}
\end{eqnarray}
$\bar{{\rm H}}$ is the charge symmetric part of the Hamiltonian and 
$\Delta {\rm H_{CSB}}$ contains Eqs.~(\ref{eq:Delta K}) 
and (\ref{eq:Delta V}). 
The first two terms of Eq.~(\ref{eq:massformula}) 
give the average mass of the nucleon 
and the third term contributes to ${\rm \Delta M}$. 
The up-down quark mass difference $\Delta m$
is determined so as to reproduce ${\rm \Delta M}$ 
by using Eq.~(\ref{eq:massformula}). \par
The second difference from the Isgur's work 
is that the Instanton Induced Interaction (III) is considered 
in this study. III has the contact spin-spin interaction and 
contributes to the difference of the masses of the Nucleon and 
$\Delta$(1232) just like 
the color magnetic interaction. We choose the coupling constant of the 
OGE, $\alpha_s$, and the III, $V_0^{(2)}$, 
so as to reproduce the nucleon-$\Delta$ mass difference in total. 
So $\alpha_s$ becomes smaller effectively by considering III. 
\subsection{\label{sec3.2}CSB in the N-N scattering}
In the calculation of the scattering lengths and analyzing powers, 
we employ the quark cluster model (QCM) \cite{oka,fae,oy,shi,osysup}, 
which describes two-nucleon systems 
in terms of their quark coordinates. The scattering wave functions, which 
are used as the unperturbated states, are calculated by solving the 
resonating group method (RGM) equation. 
By mainly technical reasons the kinetic energy term is treated 
purely in the non-relativistic way, i.e. 
the semirelativistic kinematics is not taken into account contrary to 
the case of single baryon mass. 
This approximation can be justified 
because the relativistic effect on the kinetic energy term 
is smaller for the motion of the two baryons. 
Then the kinetic energy is given as 
\begin{eqnarray}
{\rm K}&=& \sum_{i}^6 {\rm K_i} - {\rm K_G} \label{eq:K} \\
{\rm K_i} &=& (m_i+\frac{p_i^2}{2m_i}) \\
{\rm K_G}&=& \frac{{\rm P_G}^2}{2 {\rm M_G}}
\end{eqnarray}
where
\begin{eqnarray}
{\rm M_G}=\sum_i^6 m_i~~~~~{\rm P_G}=\sum_i^6 p_i
\end{eqnarray}
The RGM equation for the baryon A and baryon B is as follows, 
\begin{eqnarray}
\int \phi_{{\rm A}} (\xi_{{\rm A}}) \phi_{{\rm B}} (\xi_{{\rm B}}) 
({\rm H-E}) {\mathcal A} 
[ \phi_{{\rm A}} (\xi_{{\rm A}}) \phi_{{\rm B}} (\xi_{{\rm B}}) 
\chi(R_{{\rm AB}}) ]d\xi_{{\rm A}} d\xi_{{\rm B}} = 0 \\
\phi_A(\xi_A) = (\frac{1}{2\pi b^2})^{\frac{3}{4}} 
(\frac{2}{3\pi b^2})^{\frac{3}{4}} 
\exp(-\frac{\xi_{A1}^2}{4b^2}-\frac{\xi_{A2}^2}{3b^2})
\end{eqnarray}
$\phi_{A(B)}$ and $\xi_{A(B)}$ is the internal wave function and coordinates 
of the baryon A(B). $R_{AB}$ is the relative coordinates 
of the baryon A and B. 
The parameter $b$ is the gaussian size parameter, which represents a 
nucleon size. 
${\mathcal A}$ is the antisymmetrization operator 
for six quarks and is written as follows. 
\begin{eqnarray}
{\mathcal A}=1-{\mathcal A'}
=1-\sum_{i \in {\rm A},j \in {\rm B}} {\rm P_{ij}}
\end{eqnarray}
In the end, the following equation is obtained 
\begin{eqnarray}
[ {\rm \frac{P_{{\rm AB}}^2}{2\mu_{{\rm AB}}} }+ V_{{\rm rel}}^{(D)}(R) 
-\frac{k^2}{2\tilde{\mu}_{\rm{AB}}} ]
\chi (R)
&-&\int dR' (K^{(EX)}(R,R')+V^{(EX)}(R,R') \nonumber \\
&-&EN^{(EX)}(R,R')) \chi(R') 
= 0 \label{eq:energy}
\end{eqnarray}
where $P_{AB}$ is the momentum operator of the relative motion of 
the baryons A and B, and 
\begin{eqnarray}
E &=& {\rm \tilde{M}_A} + {\rm \tilde{M}_B} 
+ {\rm \frac{k^2}{2\tilde{\mu}_{AB}}} \\
\frac{1}{\mu_{{\rm AB}}}&=&\frac{1}{\rm{M_A}}+\frac{1}{{\rm M_B}} \\
{\rm M_{A(B)}}&=& \sum_{i \in {\rm A(B)}}^3 m_i \\
\frac{1}{\tilde{\mu}_{{\rm AB}}} &=& 
\frac{1}{{\rm \tilde{M}_A}}+\frac{1}{{\rm \tilde{M}_B}} \\
{\rm \tilde{M}_{A(B)}} &:& {\rm observed~mass~of~the~baryon~A(B)} 
\end{eqnarray}
It should be noted here that ${\rm M_{A(B)}}$ and 
${\rm \tilde{M}_{A(B)}}$ may not agree with each other completely. 
We take $m_i=313$ [MeV] in our calculation so that the difference is 
small, but for the charge symmetry breaking we assume 
that ${\rm \mu_{AB}}={\rm \tilde{\mu}_{AB}}$. 
The observed masses of the proton and neutron are given by 
\begin{eqnarray}
{\rm \tilde{M}_A} &=& \tilde{{\rm M}}
(1-\epsilon_{{\rm N}}\tau_3^{{\rm (A)}}) \\
\epsilon_{\rm N} &=& \frac{\Delta {\rm \tilde{M}}}{2\tilde{{\rm M}}} \\
\tilde{{\rm M}} &=& \frac{{\rm \tilde{M}_p}+{\rm \tilde{M}_n}}{2} 
= 939~{\rm MeV} \\
\Delta {\rm \tilde{M}} &=& {\rm \tilde{M}_n}-{\rm \tilde{M}_p} 
= 1.29~{\rm MeV} 
\end{eqnarray}
Therefore we may rewrite the kinetic energy terms as 
\begin{eqnarray}
{\rm \frac{P_{AB}^2}{2\tilde{\mu}_{AB}}}-{\rm \frac{k^2}{2\tilde{\mu}_{AB}}}
&=& {\rm \frac{P_{AB}^2-k^2}{2\tilde{\mu}} }
(1+\frac{\tau_3^{({\rm A})}+\tau_3^{({\rm B})}}{2}\epsilon_{{\rm N}}) \\
\end{eqnarray}
and the energy in Eq.~(\ref{eq:energy}) as
\begin{eqnarray}
E &=& 2\tilde{{\rm M}}
(1-\frac{\tau_3^{({\rm A})}+\tau_3^{({\rm B})}}{2}\epsilon_{{\rm N}})
+\frac{k^2}{2\tilde{\mu}}
(1+\frac{\tau_3^{({\rm A})}+\tau_3^{({\rm B})}}{2}\epsilon_{{\rm N}}) 
\nonumber \\
&=& 2\tilde{{\rm M}} + \frac{k^2}{2\tilde{\mu}} 
+ (-2\tilde{{\rm M}} + \frac{k^2}{2\tilde{\mu}})
(\frac{\tau_3^{({\rm A})}+\tau_3^{({\rm B})}}{2}\epsilon_{{\rm N}}) 
\nonumber \\
&=& {\bar E}+\Delta E_{{\rm CSB}}
\end{eqnarray}
because
\begin{eqnarray}
{\rm \frac{1}{2\tilde{\mu}_{AB}} } &=& \frac{1}{2\tilde{\mu}}
(1+\frac{\tau_3^{({\rm A})}+\tau_3^{{\rm (B)}}}{2} \epsilon_{{\rm N}}) \\
{\rm \tilde{\mu}} &=& {\rm \frac{\tilde{M}}{2}} 
\end{eqnarray}
\par
The RGM kernels $V_{{\rm rel}}^{(D)},N^{(EX)},K^{(EX)},V^{(EX)}$ 
are defined by 
\begin{eqnarray}
V_{{\rm rel}}^{(D)}(R) &=& \int {\rm d\xi_{{\rm A}}d\xi_{{\rm B}} dR_{AB}}
\phi_{{\rm A}}(\xi_{{\rm A}}) 
\phi_{{\rm B}}(\xi_{{\rm B}}) \nonumber \\
&& \sum_{i \in {\rm A} j \in {\rm B}} {\rm V}_{ij} \delta({\rm R-R_{AB}})
\phi_{{\rm A}}(\xi_{{\rm A}}) 
\phi_{{\rm B}}(\xi_{{\rm B}}) 
\\
\left(
\begin{array}{lcr}
N^{(EX)}(R',R) \\
K^{(EX)}(R',R) \\
V^{(EX)}(R',R) \\
\end{array}
\right)
&=&
\int d\xi_{{\rm A}} d\xi_{{\rm B}} d{\rm R_{AB}}
\phi_{{\rm A}}(\xi_{{\rm A}}) 
\phi_{{\rm B}}(\xi_{{\rm B}}) \delta ({\rm R'-R_{AB}})
\left(
\begin{array}{lcr}
1 \\
{\rm K} \\
{\rm V} \\
\end{array}
\right) \nonumber \\
&&{\mathcal A'}
[\delta ({\rm R-R_{AB}})
\phi_{{\rm A}}(\xi_{{\rm A}}) 
\phi_{{\rm B}}(\xi_{{\rm B}}) ] \nonumber \\
&=& 
\left(
\begin{array}{lcr}
\bar{N}^{(EX)}(R',R)  \\
\bar{K}^{(EX)}(R',R) + \Delta K_{{\rm CSB}}(R',R) \\
\bar{V}^{(EX)}(R',R) + \Delta V_{{\rm CSB}}(R',R) \\
\end{array}
\right)
\end{eqnarray}
K and V are given by Eqs.~(\ref{eq:K}) and (\ref{eq:V}) and can be divided into the charge symmetric part ${\rm \bar{K},\bar{V}}$ and the charge symmetry breaking part 
$\Delta {\rm K_{CSB}},\Delta {\rm V_{CSB}}$. Therefore RGM kernels are divided 
into the charge symmetric part $\bar{K}^{(EX)},\bar{V}^{(EX)}$ 
and the charge symmetry breaking part $\Delta K_{{\rm CSB}}^{(EX)},
\Delta V_{{\rm CSB}}^{(EX)}$. \par
In order to treat the CSB part perturbatively, we employ 
the distorted wave Born approximation (DWBA) in this study. 
we solve the following equation to obtain the distorted wave. 
\begin{eqnarray}
[ {\rm \frac{P_{{\rm AB}}^2}{2\tilde{\mu}} } 
-\frac{k^2}{2\tilde{\mu}} ]\chi_{{\rm dist}}(R)
&-&\int dR' (\bar{K}^{(EX)}(R,R')+\bar{V}^{(EX)}(R,R') \nonumber \\
&-& \bar{E}\bar{N}^{(EX)}(R,R')) \chi_{{\rm dist}}(R') 
= 0 
\end{eqnarray}
The direct kernel $V_{{\rm rel}}^{(D)}(R)$ comes from 
the electromagnetic interaction of quarks and corresponds to the 
electromagnetic interaction of baryons. 
We are interested in effects of CSB at the quark level, 
not at the hadron level. So we ignore the direct kernel. 
But we consider the exchange kernel of the electromagnetic interaction 
of quarks. 
Using the distorted wave $\chi_{{\rm dis}}(R)$, 
we estimate the following CSB parts. 
\begin{eqnarray}
{\rm (CSB~part)} &=& {\rm \frac{P_{AB}^2-k^2}{2\tilde{\mu}} }
(\frac{\tau_3^{({\rm A})}+\tau_3^{({\rm B})}}{2}\epsilon_{{\rm N}}) 
\chi_{{\rm dist}}(R)
\nonumber \\
&-& \int dR' [ \Delta K^{(EX)}_{{\rm CSB}}(R,R')
+\Delta V^{(EX)}_{{\rm CSB}}(R,R') \nonumber \\
&-&\Delta E_{{\rm CSB}} \bar{N}^{(EX)}(R,R') ]\chi_{{\rm dist}}(R') 
\label{eq:CSBpart}
\end{eqnarray}
\subsection{\label{sec3.3}CSB in the analyzing power}
There is a special CSB interaction in the neutron-proton system, 
which is called the class IV interaction, 
according to the classification by Henley and Miller \cite{text}. 
\begin{eqnarray}
V_{{\rm IV}} &\propto& (\tau_3^{{\rm A}}-\tau_3^{{\rm B}}) 
(\vec{\sigma}_{{\rm A}}-\vec{\sigma}_{{\rm B}}) \\
&&{\rm or} \nonumber \\
&& (\vec{\tau}_{{\rm A}} \times \vec{\tau}_{{\rm B}})_{{\rm z}}
(\vec{\sigma}_{{\rm A}} \times \vec{\sigma}_{{\rm B}})
\end{eqnarray}
one sees that the class IV interaction mixes spin-singlet states and 
spin-triplet states. The spin singlet-triplet mixing induces asymmetries 
of spin polarization observables such as the analyzing power. 
At the level of the quark-quark interaction, CSB in the 
spin-orbit interactions is given as [See Eqs.~(\ref{eq:LSOGE}-\ref{eq:LSEM})]
\begin{eqnarray}
{\rm V_{CSB}^{LS}} &=& {\rm V_{qSLS}^{OGE}} + {\rm V_{qALS}^{OGE}} + 
{\rm V_{qSLS}^{EM}} \label{eq:ALS} \\
%
%
{\rm V_{qSLS}^{OGE}} &=& - \sum_{i<j} (\vec{\lambda}_i \cdot \vec{\lambda}_j)
\frac{3\alpha_s \epsilon}{16\bar{m}^2}\frac{\vec{L}_{ij}}{r_{ij}^3}
[(\vec{\sigma}_i+\vec{\sigma}_j)(\tau_3^{(i)}+\tau_3^{(j)})] 
\label{eq:OGEsls} \\
{\rm V_{qALS}^{OGE}} &=& - \sum_{i<j} (\vec{\lambda}_i \cdot \vec{\lambda}_j)
\frac{\alpha_s \epsilon}{16\bar{m}^2}\frac{\vec{L}_{ij}}{r_{ij}^3}
[(\vec{\sigma}_i-\vec{\sigma}_j)(\tau_3^{(i)}-\tau_3^{(j)})] 
\label{eq:OGEals} \\
{\rm V_{qSLS}^{EM}} &=& - \sum_{i<j} 
\frac{\alpha_{em} }{16\bar{m}^2}\frac{\vec{L}_{ij}}{r_{ij}^3}
[(\vec{\sigma}_i+\vec{\sigma}_j)(\tau_3^{(i)}+\tau_3^{(j)})]
\label{eq:EMsls}
\end{eqnarray}
The first two terms of Eq.~(\ref{eq:ALS}) come from the one-gluon-exchange interaction 
and the third term from the electromagnetic interaction of quarks. 
It should be noted that the symmetric spin-orbit interaction of quarks (qSLS)
induces the class IV interaction of baryons as well as the antisymmetric 
one (qALS). Br\"{a}uer et al. calculated $\Delta A(\theta)$ 
using a similar model without including the qSLS terms \cite{brauer2}. 
They concluded that the contribution of quarks to $\Delta A(\theta)$ is 
very small. But we will see that 
the contribution of quark spin-orbit interactions, Eq.~(\ref{eq:ALS}), 
to $\Delta A(\theta)$ is large enough to reproduce the
observed $\Delta A(\theta)$. \par
Using DWBA, we calculate the following matrix elements for J=L$\le$3, 
\begin{eqnarray}
\Delta {\rm T_{CSB}} &=& 
\langle ^{3}{\rm L_J} | {\rm V_{CSB}^{LS}} | ^{1}{\rm L_J} \rangle
\end{eqnarray}
Then the total T-matrix is given as follows, 
\begin{eqnarray}
{\rm T} &=& {\rm \bar{T}_{CS}+\Delta T_{CSB}}
\end{eqnarray}
${\rm \bar{T}_{CS}}$ is obtained by solving 
the RGM equation. 
T is the regarded as a matrix based on the spin states and 
the analyzing power is given by 
\begin{eqnarray}
A_{{\rm N}}(\theta) &=& {\rm \frac{Tr[ T^{\dagger} \sigma_N T ]}
{Tr[T^{\dagger}T]}}
\end{eqnarray}
Then $\Delta A(\theta)$ is given in terms of 
${\rm \bar{T}_{CS}}$ and ${\rm \Delta T_{CSB}}$ 
\begin{eqnarray}
\Delta A(\theta) &=& A_n (\theta) -A_p (\theta) \nonumber \\
&=& {\rm \frac{2 Re Tr[\bar{T}_{CS}^{\dagger}(\sigma_n - \sigma_p)\Delta T_{CSB}]}
{Tr[\bar{T}_{CS}^{\dagger}\bar{T}_{CS}]} } \label{eq:Delta A}
\end{eqnarray}
We show the explicit forms of the T-matrix and of $\Delta A(\theta)$
in Appendix A. 
\section{\label{sec4}Results} 
The parameters in our calculation 
are determined so as to reproduce the single nucleon property. 
In order to show explicitly how much the contribution of 
the Instanton Induced Interaction (III) to the Nucleon-$\Delta$ splitting is, 
we introduce a new parameter ${\rm P_{III}}$, which denotes 
the ratio of the contribution of III 
to the whole Nucleon-$\Delta$ splitting. 
For example, when ${\rm P_{III}}=0.4$ 
the contribution of III to the Nucleon-$\Delta$ splitting is 40\%
of the whole one. 
$V_0^{(2)}$ is determined so as to reproduce the $\eta$ and $\eta'$ 
mass splitting. 
Our analysis shows that ${\rm P_{III}} \sim$ 0.4-0.5 gives the right 
$\eta$-$\eta'$ splitting. 
Here we try two values ${\rm P_{III}}=0.4$ and 0.5. 
Using the nucleon mass formula Eq.~(\ref{eq:massformula}), 
we obtain $\Delta m $ for each ${\rm P_{III}}$. 
The results are given in Table \ref{tab:table1}. 
The parameter b is the gaussian size parameter 
for the internal wave function of the nucleon, 
which represents the nucleon size. \par
%
%
%
\begin{table}[ht]
\caption{\label{tab:table1}Parameters} 
\begin{ruledtabular}
  \begin{tabular}{ccccccccc} 
  & ${\rm P_{III}}$ & $\Delta m$ & $\bar{m}$ [MeV] & b [fm] 
  & $\alpha_s$ & $a$ [  MeV/fm] & $V_{0}^{(2)}$ [MeV~fm$^3$] \\ \hline 
  A & 0.4 & 7.3 & 313 & 0.6 & 0.91 & 44.29 & -177.2 \\ 
  B & 0.5 & 5.2 & 313 & 0.6 & 0.76 & 40.34 & -221.5 \\ 
  \end{tabular}
\end{ruledtabular}
\end{table}
Another possible source of the N-$\Delta$ splitting is contribution of 
pion cloud around the baryon. 
For instance, the cloudy bag model predicts the N-$\Delta$ splitting 
of about 100 MeV \cite{CBM}. 
This effect may reduce the roles of OGE and III, but it is not taken 
into account in this approach. \par
%
%
%
%
\par
By increasing ${\rm P_{III}}$, we reduce $\alpha_s$ accordingly so that 
the N-$\Delta$ mass difference is fixed. For ${\rm P_{III}}=0.4$, 
$\alpha_s$ becomes 0.91, while $\alpha_s=1.52$ is necessary to 
reproduce the N-$\Delta$ mass difference only by OGE. 
In order to show the effect of the Instanton Induced Interaction 
to $\Delta {\rm M}$, we show contribution of each term
to $\Delta {\rm M}$ in Table \ref{tab:table2}, for various ${\rm P_{III}}$. 
The Kin,OGE and EM represent the contributions 
of the kinetic energy, the one-gluon exchange interaction 
and the electromagnetic interaction 
to $\Delta {\rm M}$. 
It should be noted that 
when ${\rm P_{III}=0}$ we can not reproduce the $\Delta {\rm M}$ 
because OGE gives large contribution, which goes to 
the opposite direction. 
This shows the essential role of the III, which reduces the OGE strength. \par
\begin{table}[ht]
\caption{\label{tab:table2}Contributions to $\Delta$M for $\Delta$ \textit{m} =6 MeV} 
\begin{ruledtabular}
\begin{tabular}{ccccc} 
${\rm P_{III}}$  & Kin & OGE & EM & ${\rm M_n-M_p}$ \\ \hline
0 & 4.72 & -5.54 & -0.41 & -1.23 \\
0.1 & 4.72 & -5.54 & -0.41 & -0.67 \\
0.2 & 4.72 & -4.99 &  -0.41  & -0.12 \\
0.3 & 4.72 & -4.43 & -0.41  & 0.44 \\
0.4 & 4.72 & -3.88 &  -0.41 & 0.99 \\
0.5 & 4.72 & -2.77 & -0.41  & 1.54 \\ 
\end{tabular}
\end{ruledtabular}
\end{table}
It is also found that the calculation of ``Strong hyperfine" for``p-n"
in Table I of Ref.~\onlinecite{Isgur} is different from our calculation 
even if we use the same potential. 
This is because Isgur considers distortion of the quark wave function 
from the u-d quark mass difference. 
However, to be consistent the distortion of the wave function 
should not contribute to the energy in the first order 
of the perturbation theory. 
\footnote{Chemtob and Yang also point out the mismatch with Isgur in their pape \cite{Yang}.} 
The contribution of the ``Strong hyperfine" to $\Delta {\rm M}$ 
should be $\frac{1}{3} \delta \frac{\Delta m}{\bar{m}}$ 
instead of $\frac{5}{24} \delta \frac{\Delta m}{\bar{m}}$ 
in Ref.~\onlinecite{Isgur}, 
where $\delta$ is the nucleon-$\Delta$ mass splitting. \par
Next we calculate $\Delta a$ using the parameters in Table \ref{tab:table1}. 
The results are shown in Table \ref{tab:table3}. 
$\bar{a}$($\bar{r}$) and $\Delta a$($\Delta r$) is the average and the difference of the scattering lengths (effective ranges) of the p-p and n-n scatterings. 
\begin{eqnarray}
\bar{a} &=& \frac{a_{pp}+a_{nn}}{2}~~~~~~~~~~
\Delta a = a_{pp}-a_{nn} \\
\bar{r} &=& \frac{r_{pp}+r_{nn}}{2}~~~~~~~~~~
\Delta r = r_{pp}-r_{nn}
\end{eqnarray}
Our results, $\Delta a=$0.79 and 0.52 [fm] for ${\rm P_{III}}=$0.4 and 0.5, 
are somewhat smaller than the observed value $\sim$ 1.5 [fm]. 
We, however, point out that $\Delta a$ is sensitive to the parameters 
because it is given by a cancellation of positive and negative terms. \par
\begin{table}[ht]
\caption{\label{tab:table3}Scattering length} 
\begin{ruledtabular}
\begin{tabular}{c|cccccc} 
& ${\rm P_{III}}$ & $\Delta m$ [MeV]& $\bar{a}$ [fm] & $\Delta a$[fm] &
$\bar{r}$[fm] & $\Delta r$[fm]  \\ \hline 
A & 0.4 & 7.3 & -17.9  & 0.79 & 2.42 & -0.39 \\ 
B & 0.5 & 5.2 & -17.9  & 0.52 & 2.46 & -0.25 \\ 
Exp\cite{review} & & & -18.1$\pm$0.5  & 1.5$\pm$0.5 & 2.80$\pm$0.12 
& 0.10$\pm$0.12 \\ 
B \cite{brauer1} &  & 5.0 & 20.07 & 0.46 & & \\ 
CY \cite{Yang} &  & 6.0 & & 2$\sim$3.5 & & \\
\end{tabular}
\end{ruledtabular}
\end{table}
In Table \ref{tab:table4}, we show each contribution of CSB terms 
Eq.~(\ref{eq:CSBpart}) to $\Delta a$. 
NMD, Kin, OGE and EM are contributions of the first term of 
Eq.~(\ref{eq:CSBpart}), 
the quark kinetic energy (including the $\Delta E_{CSB}$ term), 
the one-gluon exchange interaction and the electromagnetic interaction, 
respectively. \par
\begin{table}[ht]
\caption{\label{tab:table4}The contributions to $\Delta a$ of CSB terms [fm]} 
\begin{ruledtabular}
\begin{tabular}{c|cccc} 
 & NMD & Kin & OGE & EM \\ \hline 
A & 0.3 & -2.6 & 2.9 & 0.2  \\ 
B & 0.3 & -1.7 & 1.7 & 0.2  \\ 
\end{tabular}
\end{ruledtabular}
\end{table}
We estimate $\Delta a$ in our formulation for the parameters of 
Ref.~\onlinecite{brauer1} (B) and \onlinecite{Yang} (CY). 
The contributions of Kin and OGE should be given by 
\begin{eqnarray}
\Delta a_{{\rm Kin}} &\propto& \frac{\Delta m}{\bar{m}^2 b^2} 
\equiv \Delta b_{{\rm Kin}} \\
\Delta a_{{\rm OGE}} &\propto& \frac{\alpha_s \Delta m}{\bar{m}^3 b^3} 
\equiv \Delta b_{{\rm OGE}} 
\end{eqnarray}
In Table V we show $\Delta b_{{\rm Kin}}$ and $\Delta b_{{\rm OGE}}$ 
for the parameters of B and CY. 
Using the values of Table \ref{tab:table4} and \ref{tab:table5}, we find 
\begin{eqnarray}
\Delta a_{{\rm Kin}} + \Delta a_{{\rm OGE}} |_{{\rm B}} 
&=& -2.6 \times \frac{4.6}{8.0} + 2.9 \times \frac{5.2}{7.7} \nonumber \\
&=& -1.5 + 2.0 = 0.5 \\
\Delta a_{{\rm Kin}} + \Delta a_{{\rm OGE}} |_{{\rm C}} 
&=& -2.6 \times \frac{5.6}{8.0} + 2.9 \times \frac{8.9}{7.7} \nonumber \\
&=& -1.8 + 3.4 = 1.6 
\end{eqnarray}
These estimates suggest that our results may become larger 
by the changing the parameters. 
As $\Delta b_{{\rm Kin}}$ is larger than $\Delta b_{{\rm OGE}}$ 
in our parameter choice, the cancellation of $\Delta a_{{\rm Kin}}$ 
and $\Delta a_{{\rm OGE}}$ is stronger than the other cases. 
%
%
%
%
%
%
On the other hand $\Delta r$ is too large and has the wrong sign. 
More investigation should be done for $\Delta r$, which 
reflects not only the strength of the interaction but also 
its radial dependence. \par
\begin{figure}[ht]
\includegraphics[width=10cm,clip]{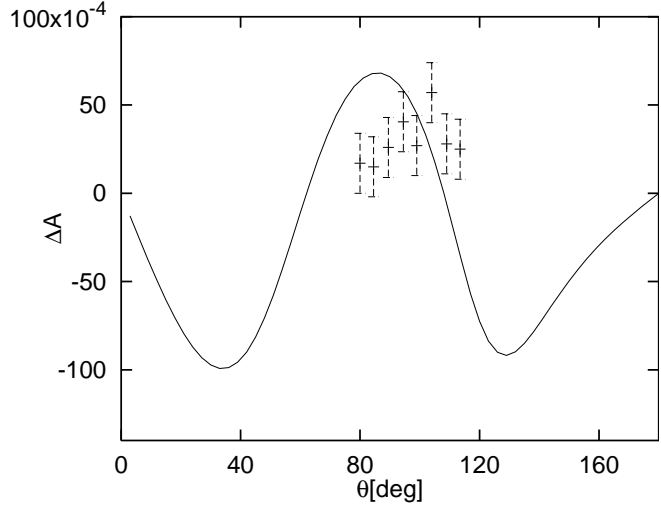}
\caption{\label{fig1}$\Delta A(\theta)$ at ${\rm E_n=183}$ [MeV].}
\end{figure}
\begin{figure}[ht]
\includegraphics[width=10cm,clip]{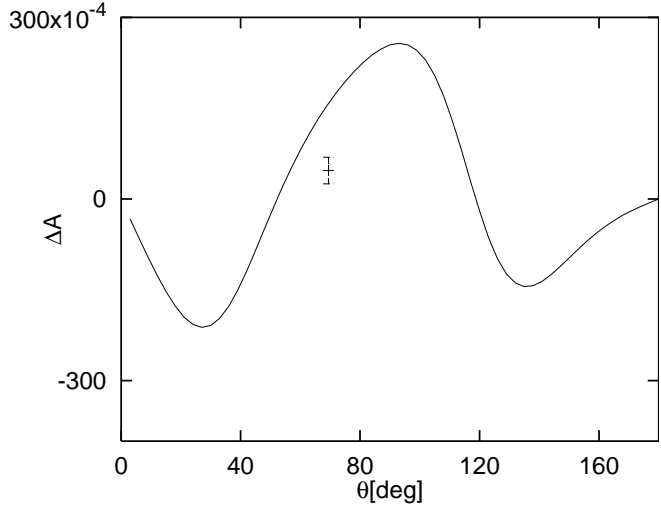}
\caption{\label{fig2}$\Delta A(\theta)$ at ${\rm E_n=477}$ MeV.} 
\end{figure}

\begin{table}[ht]
\caption{\label{tab:table5}$\Delta b_{{\rm Kin}}$ and $\Delta b_{{\rm OGE}}$} 
\begin{ruledtabular}
\begin{tabular}{c|cc} 
 & $\Delta b_{{\rm Kin}}$ [MeV] & $\Delta b_{{\rm OGE}}$ [Mev] \\ \hline 
A & 8.0 & 7.7 \\ \hline
B & 4.6 & 5.2 \\ 
CY & 5.6 & 8.9 \\ 
\end{tabular}
\end{ruledtabular}
\end{table}
Finally we calculated $\Delta A(\theta)$ at two energy points, taking 
${\rm P_{III}}=0.4$. 
The results at ${\rm E_{n}}=$ 183 and 477 [MeV]  are shown 
in Fig.~\ref{fig1} and \ref{fig2}. 
The results at ${\rm E_{n}}=$183 and 477 MeV are large 
enough to reproduce the data \cite{ALSex1,ALSex2}, 
which disagrees with the conclusion 
of Br\"{a}uer et al. \cite{brauer2}. 
The difference mainly comes from two points. 
The first point is that they consider only the antisymmetric 
spin-orbit interaction of quarks (qALS) not the symmetric 
spin-orbit interaction of quarks (sSLS). 
The factor of qSLS is three times as large as that of qALS 
(See Eqs.~(\ref{eq:OGEsls}-\ref{eq:OGEals})).
The remaining discrepancy might be attributed to their 
erroneous choice of the unit of $\gamma_{1}$ in the formula Eq.~(3.6) 
in their paper \cite{brauer2}. 
We convert their value of $\gamma_1$ in radian into that in degrees and 
obtain $\Delta A(\theta=96^{\circ})=5.4 \times 10^{-4}$, 
which is of the same order as our estimate. 
%
%
%
Our result at ${\rm E_n}=477$ [MeV] is too large. 
It is not surprising since we fit the phase shift of the N-N scattering 
up to ${\rm E_n}=400$ [MeV] and we may not apply QCM at higher energy 
and we need higher partial waves. \par
\begin{figure}[ht]
\includegraphics[width=10cm,clip]{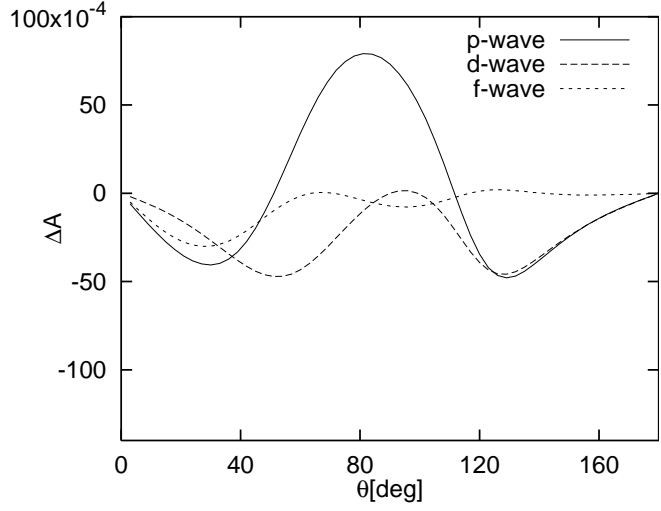}
\caption{\label{fig3}The contribution of each partial wave mixing 
at ${\rm E_n=183}$ MeV.}
\end{figure}

\begin{figure}[ht]
\includegraphics[width=10cm,clip]{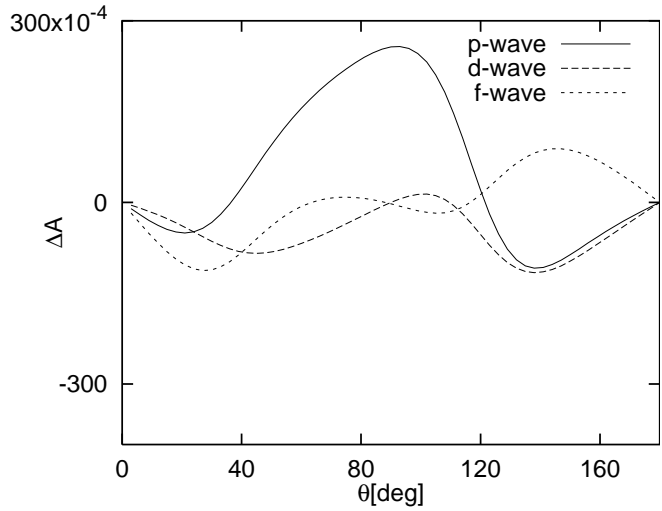}
\caption{\label{fig4}The contribution of each partial wave mixing 
at ${\rm E_n=477}$ MeV.}
\end{figure}

Fig.~\ref{fig3} and \ref{fig4} show the contributions of 
$\langle ^{1}P_{1} | \hat{T} | ^{3}P_1 \rangle $, 
$\langle ^{1}D_{2} | \hat{T} | ^{3}D_2 \rangle $ and 
$\langle ^{1}F_{3} | \hat{T} | ^{3}F_3 \rangle $ 
to $\Delta A(\theta)$. 
It is found that the contribution of 
$\langle ^{1}P_{1} | \hat{T} | ^{3}P_1 \rangle $ is dominant 
in the observed $\theta$ region. 
But the other mixings of partial wave become important 
for the other $\theta$ region. \par
\begin{figure}[ht]
\includegraphics[width=10cm,clip]{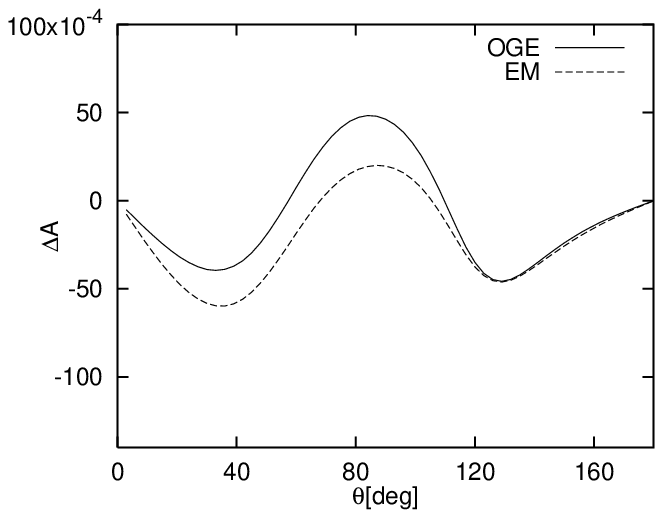}
\caption{\label{fig5}The contribution of OGE and EM 
at ${\rm E_n=183}$ MeV.}
\end{figure}

\begin{figure}[ht]
\includegraphics[width=10cm,clip]{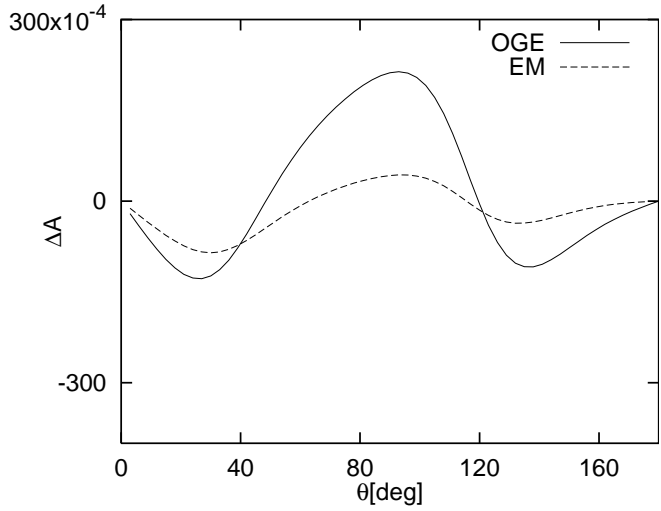}
\caption{\label{fig6}The contribution of OGE and EM 
at ${\rm E_n=477}$ MeV.}
\end{figure}

We also investigate each contribution of 
the one-gluon exchange interaction and the electromagnetic interaction. 
(Fig.~\ref{fig5} and \ref{fig6})
It is found that the contribution of OGE depends on the 
incident energy much strongly than that of the electromagnetic 
interaction does. 
This is because the dominant contribution of the EM interaction 
is the direct interaction while OGE interaction contributes 
as the exchange interaction. 
Therefore their energy dependences are different from each other, 
which may be studied by future experiment 
at various energy points. 
\section{\label{sec5}Conclusion}
We have calculated the difference of the masses of the neutron and the proton, 
${\rm \Delta M}$, the difference of the scattering lengths 
of the p-p and n-n scatterings, $\Delta a$, 
and the difference of the analyzing power 
of the proton and the neutron in the n-p scattering, $\Delta A(\theta)$, 
using the quark cluster model. 
In the calculation of ${\rm \Delta M}$, we treated the kinetic energy 
in the semirelativistic way and introduce the Instanton Induced Interaction 
(III). 
We have found that the contribution of the one-gluon-exchange interaction (OGE)
is suppressed by the introduction of the III and have determined the 
up-down quark mass difference, 
$\Delta m$=7.3 and 5.2 [MeV] for ${\rm P_{III}}=$0.4 and 0.5. \par
We have calculated $\Delta a$ for the CSB parameters fixed by ${\rm \Delta M}$. Our results are $\Delta a=$0.8 and 0.5 [fm] for ${\rm P_{III}}=$0.4 and 0.5, 
which are smaller than the observed value. 
It is found that the contribution of the u-d mass difference 
to $\Delta a$ is comparable with that from 
EM interaction because the contributions of OGE and the quark kinetic energy 
cancel out each other. 
It is pointed out that $\Delta a$ is sensitive to the choice of the quark 
model parameters because of this cancellation. \par
The P-wave CSB observable, $\Delta A(\theta)$, is calculated for 
${\rm P_{III}}=0.4$. 
It is found that 
CSB of the short range part in nuclear force is large enough to explain 
$\Delta A(\theta)$. This result is different from the conclusion of 
Br\"{a}uer et al \cite{brauer2}. 
We have found that this discrepancy is attributed to 
the introduction of the quark symmetric spin-orbit interaction and 
the erroneous choice of the $\gamma_1$ in their paper. 
We also have investigated the importance of individual mixing matrix element, 
$\langle ^{1}P_{1} | \hat{T} | ^{3}P_1 \rangle $, 
$\langle ^{1}D_{2} | \hat{T} | ^{3}D_2 \rangle $ and 
$\langle ^{1}F_{3} | \hat{T} | ^{3}F_3 \rangle $ 
and also the relative importance of the OGE and EM interaction. 
It is found that the contributions of 
$\langle ^{1}P_{1} | \hat{T} | ^{3}P_1 \rangle $ 
and OGE are dominant in the observed $\theta$ region. 
Future experiments for other angles as well as different energies 
may give us further information of the 
mixings of other partial waves and properties 
the spin-orbit parts of the OGE and 
EM interactions. 
In fact, we have observed that at ${\rm E_{n}}=477$ [MeV] 
the contributions of the higher partial waves become more important 
than at ${\rm E_{n}}=183$ [MeV]. 
The present quark model description is found to account for the 
short-range part of CSB. We would like to stress that the CSB for 
the single nucleon as well as the central and spin-orbit parts of 
the nuclear force are consistently described. 
There is a possible remaining short-range contribution 
introduced by Goldman et al. in Ref.~\onlinecite{interference} (GMS), 
which comes from interference between the QCD and QED effects. 
GMS pointed out that such an interference is necessary to explain 
the mass difference of the neutoral and charged pions. 
Its effect on the NN scattering was studied by Kao and Yang \cite{KaoYang}. 
Because this effect has much ambiguity, we have not included 
its effect in the present sutdy in order to see how the current data 
can be accounted without such complex effects. \par
Effects of longer range CSB may require further analysis. 
Approaches based on the chiral effective theory were performed 
in Refs.~\onlinecite{vanKolck}. 
Although the applicability of the chiral perturbation theory at high 
energy NN scattering phenomena is not established, its extension to 
the spin-orbit interaction might be interesting to pursue, which is 
a subject for future works. 
\appendix
\section{The decomposition of the T-matrices} 
The representations of the T-matrices in the basis of the nucleon spins 
are shown explicitly in Appendix A.
First we expand the wave function of the two nucleons as 
\begin{eqnarray}
| \vec{p},s_z^{a},s_z^{b} \rangle &=& \sqrt{4\pi} \sum_{{\rm L,S,J}}
\sum_{{\rm L_z + S_z = J_z}}^{s_z^a +s_z^b = S_z} 
\langle {\rm L,L_z,S,S_z | J ,J_z} \rangle  |{\rm ^{2S+1} L_{J}} 
\rangle {\rm Y_{L,L_z}}(\hat{p}) | s_z^a,s_z^b\rangle \label{eq:wavefun}
\end{eqnarray}
Using the wave function Eq.~(\ref{eq:wavefun}), we calculate the T-matrix. 
For example, the T-matrix of the 
$^{3}{\rm P}_0 \rightarrow ^{3}{\rm P}_0$ scattering is given by 
\begin{eqnarray}
{\mathcal T}_{{\rm ^{3}P_0} \rightarrow {\rm ^{3}P_0}}
&=& 4 \pi \sum_{m,s_z} \langle {\rm 1,m,1,s_z | 0,0} \rangle^{\ast} 
\langle {\rm 1,0,1,0 | 0,0} \rangle 
{\rm Y_{1,m}}(\hat{k})^{\ast}
{\rm Y_{1,0}}(\hat{p}) \nonumber \\
&& \langle {\rm ^{3} P_{0}} | T |{\rm ^{3} P_{0}} \rangle
\langle s_z^c,s_z^d | s_z^a, s_z^b \rangle 
|_{s_z^a +s_z^b =0,s_z^c+s_z^d =s_z} \nonumber \\
&=& \frac{1}{2} 
{\rm T}_{{\rm ^{3}P_0}} 
\left(
\begin{array}{cccc}
0 & -s ~e^{-i\phi} & -s ~ e^{-i\phi} & \\
 &  c  & c   & \\
 &  c  & c   & \\
 & s ~e^{i\phi} & s ~e^{i\phi} & 0\\
\end{array}
\right) 
\end{eqnarray}
where $\hat{p}$ is the unit vector along the initial momentum $\vec{p}$ 
and we take it along the z-axis. 
%
%
We show the the T-matrix of each partial in terms of 
$s~\equiv \sin \theta ,c \equiv \cos \theta$ and $\phi$, 
where $(\theta,\phi)$ is the scattering angle in the center of mass system. 
\begin{eqnarray}
{\mathcal T}_{{\rm ^{1}S_0} \rightarrow {\rm ^{1}S_0}}
&=& 
\frac{1}{2} {\rm  T_{^{1}S_0}} 
\left(
\begin{array}{cccc}
0 & & & \\
 & 1 & -1 & \\
 & -1 & 1 & \\
 & & & 0\\
\end{array}
\right) \\
{\mathcal T}_{{\rm ^{3}S_1}\rightarrow {\rm ^{3}S_1}} &=& 
\frac{1}{2} {\rm T_{^{3}S_1}}
\left(
\begin{array}{cccc}
2 & & & \\
 &  1&  1& \\
 &  1 & 1 & \\
 & & & 2\\
\end{array}
\right) \\
&& \nonumber \\
{\mathcal T}_{{\rm ^{1}P_1} \rightarrow {\rm ^{1}P_1}} 
&=& \frac{3}{2} {\rm T}_{{\rm ^{1}P_1}} c 
\left(
\begin{array}{cccc}
0 & & & \\
 &  1&  -1& \\
 &  -1 & 1 & \\
 & & & 0\\
\end{array}
\right) \\
&& \nonumber \\
{\mathcal T}_{{\rm ^{3}P_0} \rightarrow {\rm ^{3}P_0}} &=& \frac{1}{2} 
{\rm T}_{{\rm ^{3}P_0}} 
\left(
\begin{array}{cccc}
0 & -s ~e^{-i\phi} & -s ~ e^{-i\phi} & \\
 &  c  & c   & \\
 &  c  & c   & \\
 & s ~e^{i\phi} & s ~e^{i\phi} & 0\\
\end{array}
\right) \\
&& \nonumber \\
{\mathcal T}_{{\rm ^{3}P_1} \rightarrow {\rm ^{3}P_1}} 
&=& \frac{3}{4} {\rm T}_{{\rm ^{3}P_1}}
\left(
\begin{array}{cccc}
2c & 0 & 0 & 0 \\
s ~e^{i\phi} &  0 & 0 & -s~e^{-i\phi} \\
s ~e^{i\phi} &  0 & 0 & -s~e^{-i\phi} \\
0 & 0 & 0& 2c\\
\end{array}
\right) \\
&& \nonumber \\
{\mathcal T}_{{\rm ^{3}P_2}\rightarrow {\rm ^{3}P_2}} 
&=& \frac{1}{4} {\rm T}_{{\rm ^{3}P_2}}
\left(
\begin{array}{cccc}
6c & 2s ~e^{-i\phi} & 2s ~e^{-i\phi} & 0 \\
-3 s ~e^{i\phi} & 4c & 4c  & 
3s ~e^{-i\phi} \\
-3s ~e^{i\phi} & 4c & 4c & 
3s ~e^{-i\phi} \\
0 & -2s ~e^{i\phi} & -2s ~e^{i\phi} & 6c \\
\end{array}
\right) \\
&& \nonumber \\
{\mathcal T}_{{\rm ^{1}D_2}\rightarrow {\rm ^{1}D_2}} &=& 
\frac{5}{4} {\rm T}_{{\rm ^{1}D_2}} (3c^2-1)
\left(
\begin{array}{cccc}
0 &  &  & \\
 & 1 & -1 & \\
 & -1 & 1 & \\
 &  &  & 0 \\
\end{array}
\right) \\
&& \nonumber \\
{\mathcal T}_{{\rm ^{3}D_1} \rightarrow {\rm ^{3}D_1}} 
&=& \frac{1}{4} {\rm T}_{{\rm ^{3}D_1}}
\left(
\begin{array}{cccc}
(3c^2-1) & -6sc~e^{-i\phi} & -6sc~e^{-i\phi} & 3s^2~e^{-i2\phi} \\
3sc~e^{i\phi} & 2(3c^2-1) &  2(3c^2-1) & -3sc~e^{-i\phi} \\
3sc~e^{i\phi} & 2(3c^2-1) &  2(3c^2-1) & -3sc~e^{-i\phi}\\
3s^2~e^{2i\phi} &  6sc~e^{i\phi} & 6sc~e^{i\phi} & (3c^2-1) \\
\end{array}
\right) \\
&& \nonumber \\
{\mathcal T}_{{\rm ^{3}D_2}\rightarrow {\rm ^{3}D_2}} &=& \frac{5}{4} 
{\rm T}_{{\rm ^{3}D_2}}
\left(
\begin{array}{cccc}
(3c^2-1) &  &  & -s^2~e^{-i2\phi} \\
sc~e^{i\phi} & 0 & 0 & -sc~e^{-i\phi} \\
sc~e^{i\phi} & 0 & 0 & -sc~e^{-i\phi}\\
-s^2 ~e^{2i\phi} &  & & (3c^2-1)  \\
\end{array}
\right) \\
&& \nonumber \\
{\mathcal T}_{{\rm ^{3}D_3}\rightarrow {\rm ^{3}D_3}} &=& 
\frac{1}{4} {\rm T}_{{\rm ^{3}D_3}}
\left(
\begin{array}{cccc}
4(3c^2-1) & 6sc~e^{-i\phi} & 6sc~e^{-i\phi} & 2s^2~e^{-2i\phi} \\
-8sc~e^{i\phi} & 3(3c^2-1) & 3(3c^2-1) & 8sc~e^{-i\phi} \\
-8sc~e^{i\phi} & 3(3c^2-1) & 3(3c^2-1) & 8sc~e^{-i\phi}\\
2s^2~e^{2i\phi} & -6sc~e^{i\phi} & -6sc~e^{i\phi} & 4(3c^2-1) \\
\end{array}
\right) \\
&& \nonumber \\
{\mathcal T}_{{\rm ^{1}F_3}\rightarrow {\rm ^{1}F_3}} &=& 
\frac{7}{4} {\rm T}_{{\rm ^{1}F_3}}
(5c^3-3c)
\left(
\begin{array}{cccc}
0 &  &  & \\
 & 1 & -1 & \\
 & -1 & 1 & \\
 &  &  & 0 \\
\end{array}
\right) \\
&& \nonumber \\
{\mathcal T}_{{\rm ^{3}F_2}\rightarrow {\rm ^{3}F_2}} &=& \frac{1}{2}
{\rm T}_{{\rm ^{3}F_2}}
\left(
\begin{array}{cccc}
c(5c^2-3) & -\frac{3}{2}s(5c^2-1)~e^{-i\phi} & -\frac{3}{2}s(5c^2-1)~e^{-i\phi} 
& 5s^2 c ~e^{-2i\phi} \\
s(5c^2-1)~e^{i\phi} & \frac{3}{2}c(5c^2-3) & \frac{3}{2}c(5c^2-3) 
& -s(5c^2-1)~e^{-i\phi} \\
s(5c^2-1)~e^{i\phi} & \frac{3}{2}c(5c^2-3) & \frac{3}{2}c(5c^2-3) 
& -s(5c^2-1)~e^{-i\phi} \\
5s^2c~e^{2i\phi} & \frac{3}{2}s(5c^2-1)~e^{i\phi} & \frac{3}{2}s(5c^2-1)~e^{i\phi} & c(5c^2-3) \\
\end{array}
\right) \nonumber \\
&& \\
{\mathcal T}_{{\rm ^{3}F_3}\rightarrow {\rm ^{3}F_3}} &=&  \frac{1}{2}
{\rm T}_{{\rm ^{3}F_3}}
\left(
\begin{array}{cccc}
\frac{7}{2}c(5c^2-3) &  &  & -\frac{35}{4}s^2 c~e^{-2i\phi} \\
\frac{7}{8}s(5c^2-1)~e^{i\phi} & 0 & 0 & -\frac{7}{8}s(5c^2-1)~e^{-i\phi} \\
\frac{7}{8}s(5c^2-1)~e^{i\phi} & 0 & 0 & -\frac{7}{8}s(5c^2-1)~e^{-i\phi} \\
-\frac{35}{4}s^2 c~e^{2i\phi} &  &  & \frac{7}{2}c(5c^2-3) \\
\end{array}
\right) \\
&& \nonumber \\
{\mathcal T}_{{\rm ^{3}S_1\rightarrow ^{3} D_{1}}} &=& \frac{\sqrt{2}}{4}
{\rm T}_{{\rm ^{3}S_1 \rightarrow ^{3} D_{1}}} 
\left(
\begin{array}{cccc}
3c^2-1 & 3sc~e^{-i\phi} & 3sc~e^{-i\phi} & 3s^2~e^{-i\phi}\\
3sc~e^{i\phi} & -(3c^2-1) & -(3c^2-1) & -3sc~e^{i\phi} \\
3sc~e^{i\phi} & -(3c^2-1) & -(3c^2-1) & -3sc~e^{i\phi} \\
3s^2~e^{2i\phi} & -3sc~e^{i\phi} & -3sc~e^{i\phi} & 3c^2-1 \\
\end{array}
\right) \\
{\mathcal T}_{{\rm ^{3}D_1 \rightarrow ^{3}S_1}} &=& \frac{\sqrt{2}}{2} 
{\rm T}_{{\rm ^{3}D_1 -> ^{3}S_1}} 
\left(
\begin{array}{cccc}
1 &  &  & \\
 & -1 & -1 & \\
 & -1 & -1 & \\
 &  &  &  1 \\
\end{array}
\right) \\
{\mathcal T}_{ {\rm ^{3}P_2} \rightarrow {\rm ^{3} F_2} } &=& \frac{\sqrt{6}}{4}{\rm T}_{{\rm ^{3}P_2 \rightarrow ^{3}F_2}} 
\left(
\begin{array}{cccc}
c(5c^2-3) & s(5c^2-1)~e^{-i\phi} & s(5c^2-1)~e^{-i\phi} & 5s^2 c~e^{-2i\phi} \\
s(5c^2-1)~e^{i\phi} & -c(5c^2-3) & -c(5c^2-3)  & -s(5c^2-1)~e^{-i\phi} \\
s(5c^2-1)~e^{i\phi} & -c(5c^2-3)  & -c(5c^2-3)  & -s(5c^2-1)~e^{-i\phi} \\
5s^2 c~e^{2i\phi} & -s(5c^2-1)~e^{i\phi} & -s(5c^2-1)~e^{i\phi} & c(5c^2-3) \\
\end{array}
\right) \nonumber \\
\\
{\mathcal T}_{ {\rm ^{3} F_2} \rightarrow {\rm ^{3} P_2} } &=& \frac{\sqrt{6}}{4}
{\rm T}_{{\rm ^{3}F_2 \rightarrow ^{3}P_2}} 
\left(
\begin{array}{cccc}
2c & -s~e^{-i\phi} & -s~e^{-i\phi} & 0 \\
-s~e^{i\phi} & -2c & -2c & s~e^{-i\phi} \\
-s~e^{i\phi} & -2d & -2c & s~e^{-i\phi}\\
0 & s~e^{i\phi} & s~e^{i\phi} & 2c \\
\end{array}
\right) \\
{\mathcal T}_{{\rm ^{1}P_1} \rightarrow ^{3}P_1} &=& \frac{3\sqrt{6}}{4}
{\rm T}_{{\rm ^{1}P_1} -> ^{3}P_1} 
\left(
\begin{array}{cccc}
0 & -s~e^{-i\phi} & s~e^{-i\phi} & 0 \\
 & 0 & 0 & \\
 & 0 & 0 & \\
0 & -s~e^{i\phi} & s~e^{i\phi} & 0 \\
\end{array}
\right) \\
{\mathcal T}_{{\rm ^{3}P_1} \rightarrow ^{1}P_1} &=&  \frac{3\sqrt{6}}{4}
{\rm T}_{{\rm ^{3}P_1} \rightarrow ^{1}P_1} 
\left(
\begin{array}{cccc}
0 &  &  & 0 \\
s~e^{i\phi} & 0 & 0 & s~e^{-i\phi} \\
-s~e^{i\phi} & 0 & 0 & -s~e^{-i\phi} \\
0 &  &  & 0 \\
\end{array}
\right) \\
{\mathcal T}_{{\rm ^{1}D_2} \rightarrow ^{3}D_2} &=& \frac{5\sqrt{6}}{4}
{\rm T}_{{\rm ^{1}D_2} \rightarrow ^{3}D_2} 
\left(
\begin{array}{cccc}
0 & -sc~e^{-i\phi} & sc~e^{-i\phi} & 0 \\
 & 0 & 0 & \\
 & 0 & 0 & \\
0 & -sc~e^{i\phi} & sc~e^{i\phi} & 0 \\
\end{array}
\right) \\
{\mathcal T}_{{\rm ^{3}D_2} \rightarrow ^{1}D_2} &=& \frac{5\sqrt{6}}{4} 
{\rm T}_{{\rm ^{3}D_2} \rightarrow ^{1}D_2} 
\left(
\begin{array}{cccc}
0 &  &  &  0 \\
sc~e^{i\phi} & 0 & 0 & sc~e^{-i\phi} \\
-sc~e^{i\phi} & 0 & 0 & -sc~e^{-i\phi} \\
0 &  &  & 0 \\
\end{array}
\right) \\
{\mathcal T}_{{\rm ^{1}F_3} \rightarrow ^{3}F_3} &=& \frac{7\sqrt{3}}{8}
{\rm T}_{{\rm ^{1}F_3} \rightarrow ^{3}F_3} 
\left(
\begin{array}{cccc}
0 & -s*(5c^2-1)~e^{-i\phi} & s*(5c^2-1)~e^{-i\phi} & 0 \\
 & 0 & 0 & \\
 & 0 & 0 & \\
0 & -s*(5c^2-1)~e^{i\phi} & s*(5c^2-1)~e^{i\phi} & 0 \\
\end{array}
\right) \\
{\mathcal T}_{{\rm ^{3}F_3} \rightarrow ^{1}F_3} &=& \frac{7\sqrt{3}}{8}
{\rm T}_{{\rm ^{3}F_3} \rightarrow ^{1}F_3} 
\left(
\begin{array}{cccc}
0 &  &  & 0 \\
s(5c^2-1)~e^{i\phi} & 0 & 0 & s(5c^2-1)~e^{-i\phi} \\
-s(5c^2-1)~e^{i\phi} & 0 & 0 & -s(5c^2-1)~e^{-i\phi} \\
0 &  &  & 0 \\
\end{array}
\right) 
\end{eqnarray}
Substituting the above T-matrices into the denominator and 
the numerator of Eq.~(\ref{eq:Delta A}), we obtain 
%
\begin{eqnarray}
{\rm Tr}[{\rm \bar{T}^{\dagger}\bar{T}}] &=& 
\frac{1}{8} | -2 {\rm T_{^{1}S_0}}+2{\rm T_{^{3}S_1}}
-2\sqrt{2}{\rm T_{^{3}S_1 \rightarrow ^{3}D_1}}\nonumber \\
&+&
2c( -3{\rm T_{^{1}P_1}}
+{\rm T_{^{3}P_0}}+2{\rm T_{^{3}P_2}}
-\sqrt{6}{\rm T_{^{3}P_2 \rightarrow ^{3}F_2}}) \nonumber \\
&+& (3c^2-1)
(-5{\rm T_{^{1}D_2}}+2{\rm T_{^{3}D_1}}+3{\rm T_{^{3}D_3}}
-\sqrt{2}{\rm T_{^{3}S_1 \rightarrow ^{3}D_1}}) \nonumber \\
&+&c(5c^2-3)(-7{\rm T_{^{1}F_3}}+3{\rm T_{^{3}F_2}}
-\sqrt{6}{\rm T_{^{3}P_2 \rightarrow ^{3}F_2}}) |^2 \nonumber \\
&+& 
\frac{1}{8} | 2 {\rm T_{^{1}S_0}}+2{\rm T_{^{3}S_1}}
-2\sqrt{2}{\rm T_{^{3}S_1 \rightarrow ^{3}D_1}} \nonumber \\
&+&
2c( 3{\rm T_{^{1}P_1}}
+{\rm T_{^{3}P_0}}+2{\rm T_{^{3}P_2}}
-\sqrt{6}{\rm T_{^{3}P_2 \rightarrow ^{3}F_2}}) \nonumber \\
&+& (3c^2-1)
(5{\rm T_{^{1}D_2}}+2{\rm T_{^{3}D_1}}+3{\rm T_{^{3}D_3}}
-\sqrt{2}{\rm T_{^{3}S_1 \rightarrow ^{3}D_1}}) \nonumber \\
&+&c(5c^2-3)(7{\rm T_{^{1}F_3}}+3{\rm T_{^{3}F_2}}
-\sqrt{6}{\rm T_{^{3}P_2 \rightarrow ^{3}F_2}}) |^2 \nonumber \\
&+& 
\frac{1}{8} | 4 {\rm T_{^{3}S_1}} + 2 \sqrt{2}
{\rm T_{^{3}S_1 \rightarrow ^{3}D_1}} +
2c (3{\rm T_{^{3}P_1}}+ 3{\rm T_{^{3}P_2}}
+ \sqrt{6}{\rm T_{^{3}P_2 \rightarrow ^{3}F_2}}) \nonumber \\
&+& (3c^2-1)({\rm T_{^{3}D_1}}+5{\rm T_{^{3}D_2}}+4{\rm T_{^{3}D_3}} +
\sqrt{2} {\rm T_{^{3}S_1 \rightarrow ^{3}D_1}}) \nonumber \\
&+& c(5c^2-3) (2{\rm T_{^{3}F_2}}+7{\rm T_{^{3}F_3}}
+\sqrt{6} {\rm T_{^{3}P_2 \rightarrow ^{3}F_2}}) 
|^2 \nonumber \\
&+&
\frac{1}{8} s^4 | 3{\rm T_{^{3}D_1}} -5 {\rm T_{^{3}D_2}} +2 {\rm T_{^{3}D_3}}
+3\sqrt{2} {\rm T_{^{3}S_1 \rightarrow ^{3}D_1}} \nonumber \\
&+&c(10 {\rm T_{^{3}F_2}}-\frac{35}{2} {\rm T_{^{3}F_3}}
+5\sqrt{6} {\rm T_{^{3}P_2 \rightarrow ^{3}F_2}}
)
|^2 \nonumber \\
&+& \frac{1}{4} s^2 | 2 {\rm T_{^{3}P_0}} -2 {\rm T_{^{3}P_2}} + 
\sqrt{6}{\rm T_{^{3}P_2 \rightarrow ^{3}F_2}} \nonumber \\
&+& 3c (2 {\rm T_{^{3}D_1}} -2 {\rm T_{^{3}D_3}}
-\sqrt{2}{\rm T_{^{3}S_1 \rightarrow ^{3}D_1}}) \nonumber \\
&+& (5c^2-1) (3{\rm T_{^{3}F_2}} 
- \sqrt{6} {\rm T_{^{3}P_2 \rightarrow ^{3}F_2}})
|^2 \nonumber \\
&+& \frac{1}{4} s^2 | 3{\rm T_{^{3}P_1}} -3 {\rm T_{^{3}P_2}} - 
\sqrt{6} {\rm T_{^{3}P_2 \rightarrow ^{3}F_2}}  \nonumber \\
&+&c(3{\rm T_{^{3}D_1}}+5{\rm T_{^{3}D_2}}-8{\rm T_{^{3}D_3}}
+3\sqrt{2}{\rm T_{^{3}S_1 \rightarrow ^{3}D_1}}) \nonumber \\
&+& (5c^2-1)(2{\rm T_{^{3}F_2}}+\frac{7}{4}{\rm T_{^{3}F_3}}
+\sqrt{6} {\rm T_{^{3}P_2 \rightarrow ^{3}F_2}})
|^2
\\
{\rm Tr}[{\rm \bar{T}^{\dagger} (\sigma_n -\sigma_p) \Delta T_{CSB}}] &=& -\frac{1}{4}i
(3\sqrt{6}s{\rm T_{^{3}P_1 \rightarrow ^{1}P_1}} 
+5\sqrt{6}sc{\rm T_{^{3}D_2 \rightarrow ^{1}D_2}}
+\frac{7\sqrt{3}}{2}s(5c^2-1) {\rm T_{^{3}F_3 \rightarrow ^{1}F_3}}) \nonumber \\
&&
\{
4{\rm T_{^{1}S_0}}+4{\rm T_{^{3}S_1}}
+2\sqrt{2}{\rm T_{^{3}S_1 \rightarrow ^{3}D_1}} \nonumber \\
&+&2c(3{\rm T_{^{3}P_1}}+3{\rm T_{^{3}P_2}}
+\sqrt{6} {\rm T_{^{3}P_2 \rightarrow ^{3}F_2}}) \nonumber \\
&+& (3c^2-1)(10{\rm T_{^{1}D_2}}+{\rm T_{^{3}D_1}}+5{\rm T_{^{3}D_2}}
+4{\rm T_{^{3}D_3}}+3\sqrt{2}{\rm T_{^{3}S_1 \rightarrow ^{3}D_1}} \nonumber \\
&+&(5c^2-3)c(14{\rm T_{^{1}F_3}}+2{\rm T_{^{3}F_2}}+7{\rm T_{^{3}F_3}}
+\sqrt{6} {\rm T_{^{3}P_2 \rightarrow ^{3}F_2}}) \nonumber \\
&+&s^2(3{\rm T_{^{3}D_1}}-5{\rm T_{^{3}D_2}}+2{\rm T_{^{3}D_3}}
+3\sqrt{2}{\rm T_{^{3}S_1 \rightarrow ^{3}D_1}}) \nonumber \\
&+& s^2 c ( 10{\rm T_{^{3}F_2}}-\frac{35}{2} {\rm T_{^{3}F_3}} 
+5\sqrt{6} {\rm T_{^{3}P_2 \rightarrow ^{3}F_2}})
\}
\end{eqnarray}
\end{document}